\documentclass{emulateapj}

 \usepackage{xcolor}
\usepackage{hyperref}
\hypersetup{
    colorlinks = true,
    citecolor = [cmyk]{1 .68 0 .12},
    urlcolor = [cmyk]{0 .95 1 0},
    linkcolor = [cmyk]{0 .95 1 0}
}
\usepackage[super]{nth} 
\usepackage{multirow, rotating}

\newcommand{\angst}{$\textrm{\AA}$}

\newcommand{\rp}{\textrm{R}_\textrm{p}/\textrm{R}_\star}
\newcommand{\df}{\Delta\textrm{F}_{\rm max}}
\slugcomment{Accepted for publication in ApJ}
 
\shorttitle{Stellar activity impact on X-ray and UV transits}
\shortauthors{Llama \& Shkolnik}

\begin{document}  
 

\title{Transiting the Sun: The impact of stellar activity on X-ray and ultraviolet transits}   


\author{J. Llama and E.~L. Shkolnik}
\affil{Lowell Observatory, 1400 W Mars Hill Rd,
    Flagstaff, Arizona 86001}
\email{Email: joe.llama@lowell.edu}



 
\begin{abstract}
Transits of hot Jupiters in X-rays and the ultraviolet have been shown to be both deeper and more variable than the corresponding optical transits. This variability has been attributed to hot Jupiters having extended atmospheres at these wavelengths. Using resolved images of the Sun from NASA's Solar Dynamics Observatory spanning 3.5 years of Solar Cycle 24 we simulate transit light curves of a hot Jupiter to investigate the impact of Solar like activity on our ability to reliably recover properties of the planet's atmosphere in soft X-rays (94 \angst), the UV (131-1700 \angst), and the optical (4500 \angst).  We find that for stars with similar activity levels to the Sun, the impact of stellar activity results in the derived radius of the planet in soft X-ray/EUV to be underestimated by up-to 25\% or overestimated by up-to 50\% depending on whether the planet occults active regions. We also find that in up-to 70\% of the X-ray light curves the planet transits over bright star spots. In the far ultraviolet (1600 \& 1700 \angst), we find the mean recovered value of $\rp$ to be over-estimated by up-to 20\%. For optical transits we are able to consistently recover the correct planetary radius. We also address the implications of our results for transits of WASP-12b and HD 189733b at short wavelengths. 
 
\end{abstract} 

\section{Introduction}
Understanding the impact of stellar activity on transit and radial velocity data is not only fundamental for finding Earth-sized planets but also in determining the composition of both the largest and smallest exoplanets. For instance, by observing transits at multiple wavelengths the composition of the atmosphere of the planet can be inferred through transmission spectroscopy (see for example \citealt{Sing:2011dn,Nikolov:2013ig,Stevenson:2014hp,Gibson:2012ip}). This process requires an accurate and precise measurement of the planetary to stellar radius ratio ($\rp$) at various wavelengths.  It is therefore crucial that the effects of stellar activity on transit light curves are well understood since activity will alter the shape of the transit light curve. 
 
The impact of stellar activity on transit light curves at different wavelengths has been used to not only learn about the planet but also the host star itself. At optical wavelengths, bumps in the transit light curve caused by the planet occulting star spots can be used to infer the presence of magnetically active regions on the stellar surface. By tracking the locations of these bumps in consecutive transits the spin-orbit-alignment between the orbital plane of the planet and rotation axis of the star can be derived and this value is in good agreement with Rossiter-Mclaughlin measurements \citep{SanchisOjeda:2011hd,Winn:2010da}. For misaligned systems it may be possible to track changes in the location of bumps in the light curve over time to recover stellar butterfly patterns to investigate whether other stars exhibit such cyclic evolution in the distribution of star spots \citep{Llama:2012jl,SanchisOjeda:2013iv}. Understanding the properties of stellar butterfly patterns will provide insight into the stellar dynamo \citep{Berdyugina:2005wb}.

\begin{deluxetable*}{llllll}
\centering
\tablecaption{Wavelengths observed by the Atmospheric Imaging Assembly (AIA) on-board NASA's Solar Dynamics Observatory (SDO). Adapted from \url{http://aia.lmsal.com/public/instrument.htm}.}
\startdata
\hline
Wavelength & Regime & Primary ions & Source region& Temperature (K)& Limb  \\
(\AA)&  &  &  &   & bright/dark  \\
\hline
94 & Soft X-ray / EUV & Fe XVIII & Flaring and quiet corona& $6.3\times10^6$ & Brightened \\
131 & EUV & Fe VIII, XX, XXII & Flaring and quiet corona & $4\times10^5$, $10^7$, $1.6\times10^7$ & Brightened \\
171 & EUV & Fe IX & Quiet corona and upper transition region & $6.3\times10^5$ & Brightened \\
193 & EUV & Fe XII, XXIV & Corona and hot flare plasma & $1.2\times10^6$, $2\times10^7$ & Brightened \\
211 & EUV & Fe XIV & Active regions of the corona & $2\times10^6$ & Brightened \\
304 & EUV & He II & Chromosphere and transition region & $5\times10^4$ & Brightened \\
335 & EUV & Fe XVI & Active regions of the corona & $2.5\times10^6$ & Brightened \\
1600 & FUV & C IV, continuum & Transition region and upper photosphere & $10^5$, 5$\times10^3$ & Darkened \\
1700 & FUV & continuum & Temperature minimum, photosphere & $5\times10^3$ & Darkened \\
4500 & Optical & continuum & Photosphere & $5\times10^3$ & Darkened
\enddata
\label{tab:aia} 
\end{deluxetable*} 
 
Asymmetries in the near-UV (2800\angst) light curve of WASP-12b observed using \textit{Hubble Space Telescope (HST)} have revealed an early-ingress and also a deeper transit when compared to the optical light curve \citep{Fossati:2010do,Haswell:2012ft}. However, the end of the near-UV transit coincided with the optical egress implying the presence of additional, asymmetrically distributed absorbing material in the exosphere of the planet in the near-UV.

One potential explanation for a deeper, asymmetrical transit is that heavily inflated hot Jupiters such as WASP-12b can overflow their Roche lobes (\citealt{Gu:2003hr,Li:2010fi,Ibgui:2010ev}) which may result in an optically thick accretion stream from the planet onto the star \citep{Lai:2010he}. Another solution is that the early ingress is the signature of the stellar wind impacting on the magnetosphere of the planet creating a magnetospheric bow shock \citep{Vidotto:2010jha}. This idea was modeled using Monte-Carlo radiative simulations in \citet{Llama:2011de} where they were able to fit the near-UV observations of \citet{Fossati:2010do} with a bow shock model that suggests the planet's magnetosphere extends out to a distance of $\sim 5 \textrm{R}_\textrm{P}$. 
  
Hot Jupiters orbiting much closer to their parent star are exposed to higher levels of stellar irradiance than the gas giants in our Solar System. This increase in stellar irradiance may heat the exosphere of the planet leading to mass loss (see for example \citealt{Lammer:2003eh,LecavelierDesEtangs:2004jm}). It may be possible to detect such signatures in temporal variations of transit light curve depth and duration.
 
Observations of HD 189733b at multiple wavelengths have been taken searching for evidence of an inflated planetary atmosphere and mass loss from the planet. \citet{BenJaffel:2013ei} reported a potential early ingress and also a deeper transit depth of the C \textsc{ii} emission line at 1335 \angst\, which they attribute to the magnetospheric bow shock occulting star light. Their MHD modeling predicted the bow shock will have a stand-off distance of $\sim 17\,\textrm{R}_\textrm{P}$. 

Lyman-$\alpha$ transits have been used to study both atmospheric escape and inflated planetary atmospheres on a number of hot-Jupiters. \citet{VidalMadjar:2003bl} observed transits in Lyman-$\alpha$ (1215 \angst) of HD 209458b and found the first evidence of an extended exoplanetary atmosphere with a Lyman-$\alpha$ transit depth $\sim10\times$ deeper than the corresponding optical depth. \citet{Linsky:2010fo} also observed HD 209458b and found a deeper depth in their C \textsc{ii} and Si \textsc{iii} transit light curves, suggesting mass loss from the planet. \citet{LecavelierDesEtangs:2012jq} observed transits of HD 189733b using \textit{HST/STIS} in Lyman-$\alpha$ at two epochs. In 2011 September they found an additional 14.4\% absorption and also hints of an early ingress when compared to their observations in 2010 April, which were similar to the optical light curve. They attribute this variability to the stellar wind directly influencing the evaporation rate of the upper atmosphere of the planet. Observations of 55 Cancri b have shown a partial eclipse in Lyman-$\alpha$ that coincides with the predicted transit time, suggesting the planet has an extended atmosphere that grazes the stellar disk \citep{Ehrenreich:2012bq}. \citet{Kulow:2014bv} observed the Hot-Neptune GJ 436b in Lyman-$\alpha$ and found not only a deeper transit, but also that the deepest part of the Lyman-$\alpha$ transit occurred after the egress of the optical transit, suggesting this planet has an extended Hydrogen tail. 

\citet{Llama:2013il} investigated the impact of the varying stellar wind conditions experienced by HD 189733b as it orbits around its parent star by modeling the properties of the stellar wind from the surface of the star to the orbital distance of the planet using observed magnetic maps of the star obtained by \citet{Fares:2010hq}. They found that variability in the density and pressure of the stellar wind will alter the orientation and density of the resultant bow shock, which in turn caused the predicted near-UV light curves of HD 189733b to be highly variable in both depth and duration.

 \begin{figure} 
   \centering
   \includegraphics[width=3.5in]{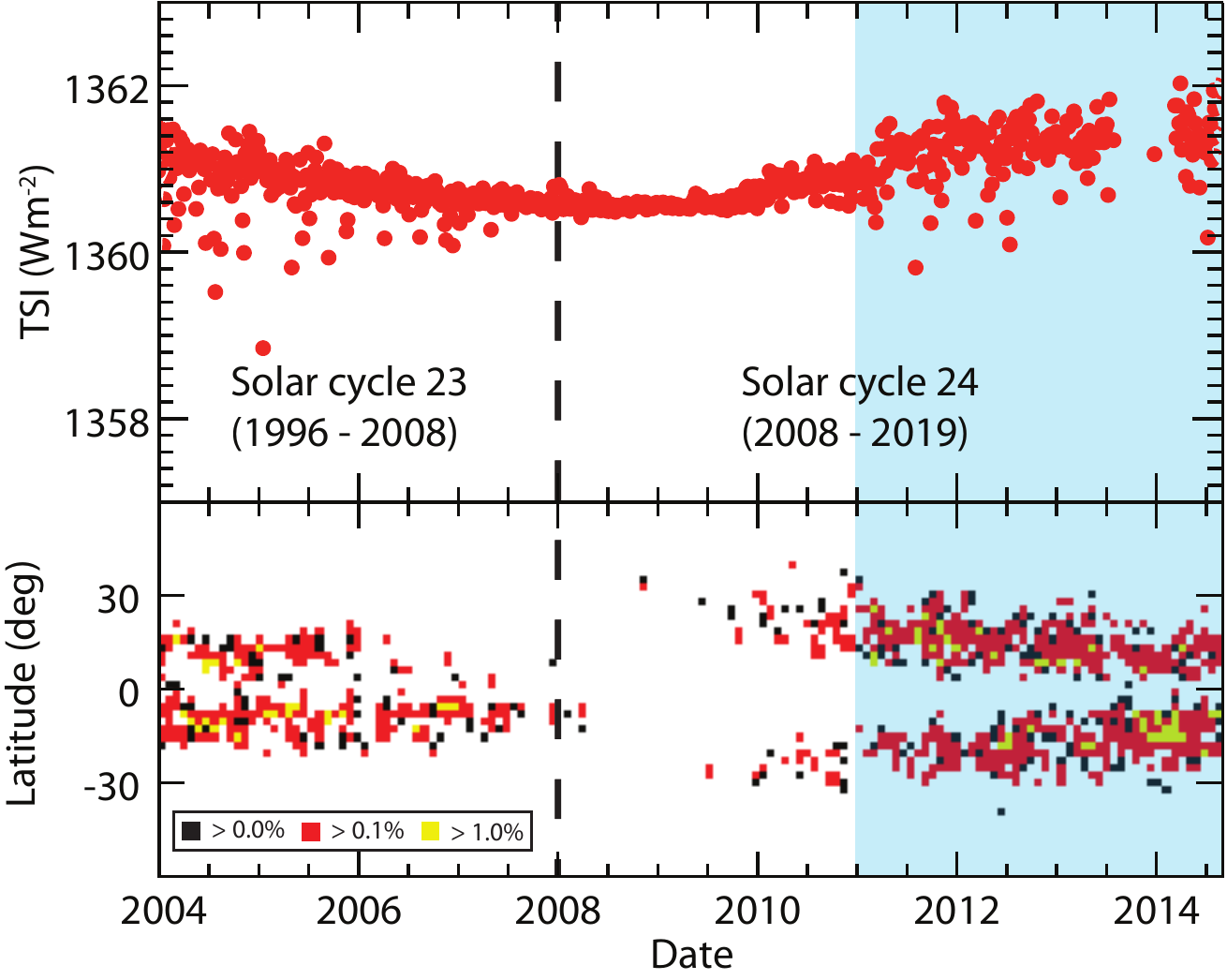} 
   \caption{\textit{Top:} Total Solar Irradiance (TSI) data from NASA's VIRGO instrument on the SOHO instrument for the past ten years, covering Solar cycle 23 and 24 \citep{Frohlich:1988hg}. \textit{Bottom:} The Solar butterfly pattern showing the fraction of the Solar disk covered in sunspots, produced using the data from \citet{Hathaway:2010hz}. The blue region shows the timespan for this investigation. During this time the Sun was moving from activity minimum toward maximum, and the active belts have migrated from $\pm 35^\circ$ to $\pm5^\circ$.}
   \label{fig:tsi}
\end{figure} 
\begin{figure*} 
\centering  
\includegraphics[width=7in]{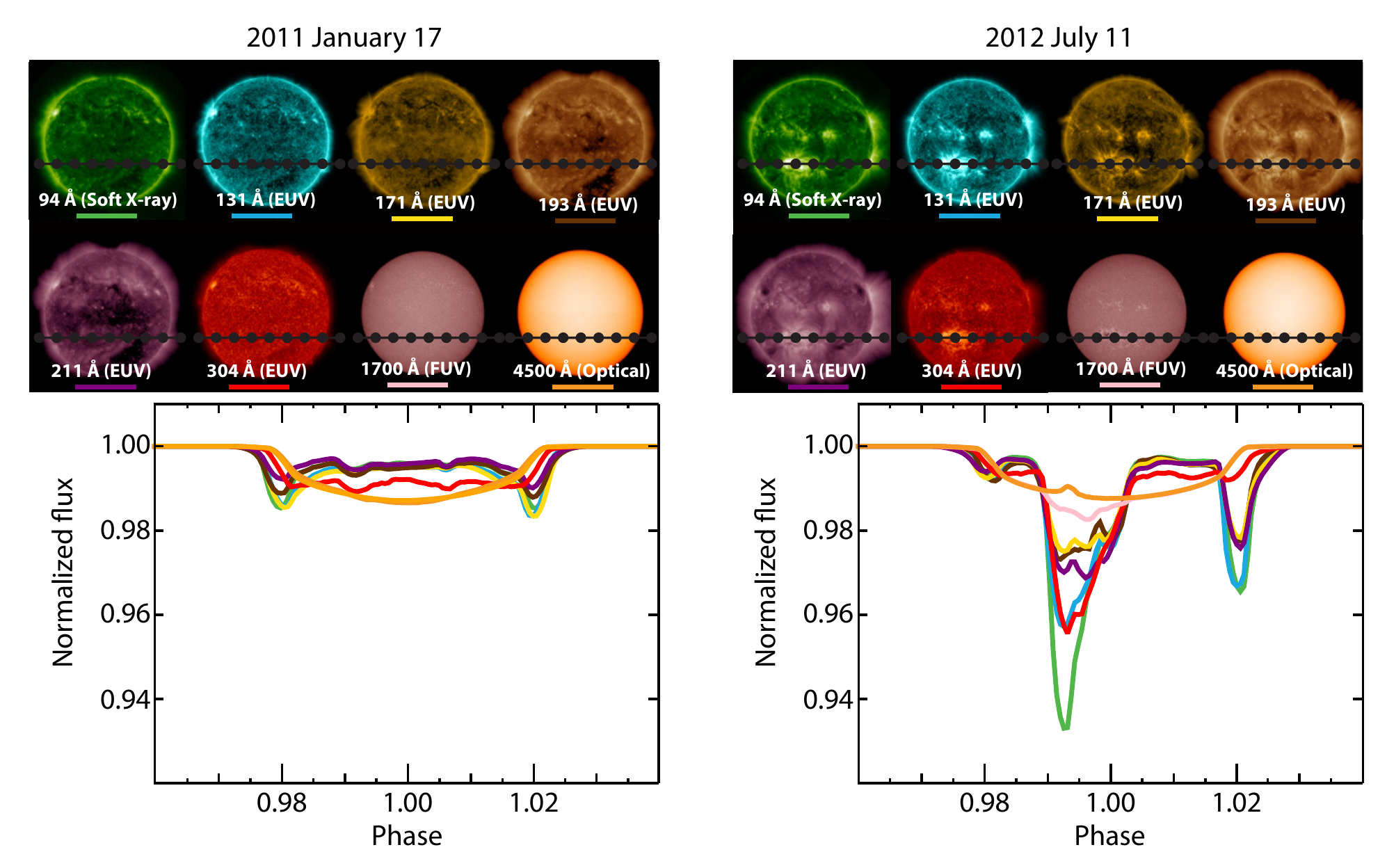}  
\caption{ {\it Top:} Two sets of simultaneous images of the Sun obtained from NASA's SDO spacecraft. Over plotted on each image is the trajectory of a simulated hot Jupiter with $\rp = 0.1$ and $\textrm{b}=-0.3$. The left panel shows images where there was little activity along the transit chord, whilst the right panel is an example of when the transit chord intersects active regions. {\it Bottom:} The resultant transit light curves are color coded by wavelength. At soft X-ray/EUV wavelengths the Sun is limb-brightened and so the transit is much deeper at ingress and egress than mid-transit. The optical transits that intersect active regions (right) show a bump in the light curve. At shorter wavelengths where active regions appear bright, the light curves exhibit a dip as the planet occults an active region.}
\label{fig:transits}
\end{figure*} 

\citet{Poppenhaeger:2013wx} observed multiple X-ray (1\angst\, - 63\angst) transits of HD 189733b with \textit{Chandra} and found a combined transit depth of 6 - 8\%, which is $\sim3\times$ deeper than the depth in the optical of 2.4\%. They attribute this additional absorption to the planet having an extended atmosphere that is opaque to X-rays but transparent to optical light. They note however, that given the difficulty of observing transits in X-rays it is not possible to rule out the increased transit depth being caused by stellar activity without further observations. Figure 9 of \citet{Poppenhaeger:2013wx} shows how the derived transit depth from all the possible combinations of their X-ray light curves varies between 2.3\% - 9.4\%.

In this paper we use resolved images of the Sun in soft X-rays, the extreme ultraviolet (EUV), the far ultraviolet (FUV), and the optical to simulate transit light curves to investigate the extent to which stellar activity impacts light curves at these wavelengths and to determine how reliably we can recover properties of a transiting hot Jupiter.

\section{Solar observations and transit model}
Observations of the Sun present a unique opportunity to investigate the effects of stellar activity on exoplanet transit light curves. NASA's Solar Dynamics Observatory (SDO) has been observing the Sun since early 2010. The Atmospheric Imaging Assembly (AIA) on-board SDO images the solar disk in ten wavelengths every ten seconds spanning the soft X-ray/EUV through to the optical (with a spatial resolution 1\arcsec\, and a CCD of 4096 px x 4096 px). A summary of the wavelengths observed by AIA with their emission is listed in Table \ref{tab:aia}. We note that it is not currently possible to observe stars other than the Sun in the EUV; however, we include the findings of these wavelengths for completeness.
 
Using the Virtual Solar Observatory {\sc idl} routines provided by SolarSoft\footnote{\url{http://www.lmsal.com/solarsoft/}}, we downloaded AIA data of the Sun at all ten wavelengths. We acquired images at a cadence of one image every 24 hours between 2011 January 01 and 2014 September 01. In total we obtained $\sim1300$ images at each of the ten wavelengths. We checked each image for saturated pixels using the {\sc NSATPIX} keyword in the FITS header of each image and found the fraction of saturated pixels in the images to be negligible. The most heavily affected image had a saturated area $<0.03\%$ meaning that saturation will not affect our simulated light curves. 

The top panel of Figure \ref{fig:tsi} shows the composite Total Solar Irradiance (TSI) between 2004 -- 2014 \citep{Frohlich:1988hg,Frohlich:2000tv}. The bottom panel shows the Solar butterfly pattern during this time \citep{Hathaway:2010hz}. The shaded blue region identifies the timespan of this investigation and demonstrates that between 2011 January and 2014 September the Sun has moved from activity minimum toward activity maximum and the active belts migrated from $\pm 35^\circ$ toward the equator.

The top panel of Figure \ref{fig:transits} shows simultaneous images of the Solar disk from AIA in eight wavelengths, namely, soft X-ray/EUV (94 \angst), EUV (131, 171, 193, 211, \& 304 \angst), the FUV (1700 \AA), and the optical (4500 \AA) on two separate dates (left and right panels)\footnote{For brevity, we do not show 335 \angst\, and 1600 \angst\, as these results are similar to 304 \angst\, and 1700 \angst\, respectively.}. The optical and FUV images demonstrate how the Sun appears relatively inactive, with only a few active regions present on the surface. At short wavelengths the Sun appears more active than when viewed in the optical. The EUV/FUV images reveal how relatively small sun spots in the optical correspond to very bright, large, extended regions at short wavelengths. It is this increase in activity and active region coverage at shorter wavelengths that will have the greatest effect on exoplanet transit light curves.

We simulated the transit light curve of an exoplanet over each image of the Solar disk. To compute the light curves we adapted the transit code developed in \citet{Llama:2011de} and \citet{Llama:2013il}. This code was originally created to simulate light curves of a planet transiting over a limb-darkened stellar disk. Rather than simulating a stellar disk, we were able to use the images from SDO. We assumed the planet to be a hot Jupiter with $\rp = 0.1$. To enable a comparison between the light curves at each wavelength we assumed the radius of the planet to be the same in all the simulations, i.e., the size of the planet's atmosphere does not change with wavelength. 

On the Sun, the activity belts are confined to within $\pm 35^\circ$ of the equator. We therefore chose two values for the impact parameter, $b$. Firstly, $b=0$, so that  the planet transits across the equator of the star and the transit chord will not intersect the activity belts and $b=-0.3$ so the planet occasionally transits over active regions. 

Finally, we assumed the planet to be completely dark, i.e., any region of the stellar disk occulted by the planet will have a flux of zero. We assumed a high cadence transit observation such that each light curve is comprised of one hundred data points, and the semi major axis of the planet to be $a\sim0.02$ au so that the planet transits once every 24 hours. As the duration of the transit event is much shorter than the rotation period of the Sun, we assumed the Sun to be static in each of our simulations. At each step, the location of the planet over the stellar disk was computed and the loss in flux caused by the planet occulting star light was calculated to produce the simulated light curve. The effects of lower (and more realistic) cadence observations are discussed in Section \ref{sec:discussion}.

\section{Results}

\subsection{Impact of stellar activity as a function of wavelength}
 
The top panel of Figure \ref{fig:transits} shows sample transit paths of a hot Jupiter with $\rp = 0.1$ and $b=-0.3$ across the Solar disk in multiple wavelengths. The images reveal that in the soft X-ray and EUV regime, the stellar disk is limb-brightened rather than limb-darkened. This difference in the distribution of star light will change the shape of the transit light curve. The final column of Table \ref{tab:aia} lists whether the Sun is limb-brightened or limb-darkened for each of the observed wavelengths. The images in the left panel of Figure \ref{fig:transits} are examples of when there is little activity along the transit path. As such, the simulated transit light curves will be examples of nearly quiescent transits. The images in the right panel of Figure \ref{fig:transits} show active regions on the disk of the Sun intersecting the transit path causing the transit light curves to deviate from the nearly quiescent shape. 

The bottom panel of Figure \ref{fig:transits} shows the corresponding simulated transit light curves (color coded by wavelength). The left panel shows nearly quiescent light curves where the planet has not transited over large active regions. In these high energy transits, a dip during both ingress and egress is registered in the light curve. The transit is also shallower at mid-transit when compared to the optical light curve. This is a consequence of the stellar disk being limb-brightened at these wavelengths, so the planet occults a much larger proportion of the total star light as it transits over the limbs of the star, as opposed to the optical case where the light is more evenly distributed over the entire disk of the star.  
 
At short wavelengths, active regions have a higher contrast ratio compared to their surroundings than sunspots in the optical. Therefore, at these wavelengths, when the planet transits over an active region, the amount of light blocked by the planet becomes greater which produces a negative dip in the light curve. There have been efforts in the literature to simulate the effects bright spots occultations and the corresponding dips in light curves have on exoplanet transmission spectroscopy \citep{Oshagh:2014hq}. The light curves in the right panel of Figure \ref{fig:transits} show the effects of the planet transiting over active regions on the transit shape.  In the optical, star spots are cooler than their surroundings and so appear dark. Therefore, when the planet transits over a star spot, the fractional loss in light is less and a positive bump is registered in the light curve (e.g., \citealt{Rabus:2009boa}). 

The evolution of unocculted active regions on the stellar disk make it difficult to accurately normalize the stellar flux level. Throughout this work we have assumed the Sun to be static during a transit observation. Given the relatively low activity level, and slow rotation period of the Sun, this is a reasonable assumption. It is worthy of note however, that for more active stars the corona will be evolving rapidly and obtaining a long baseline is crucial in accurately normalizing the data and recovering an accurate value of $\rp$.

\subsection{Transit asymmetries}
 \begin{figure*} 
   \centering
   \includegraphics[width=7in]{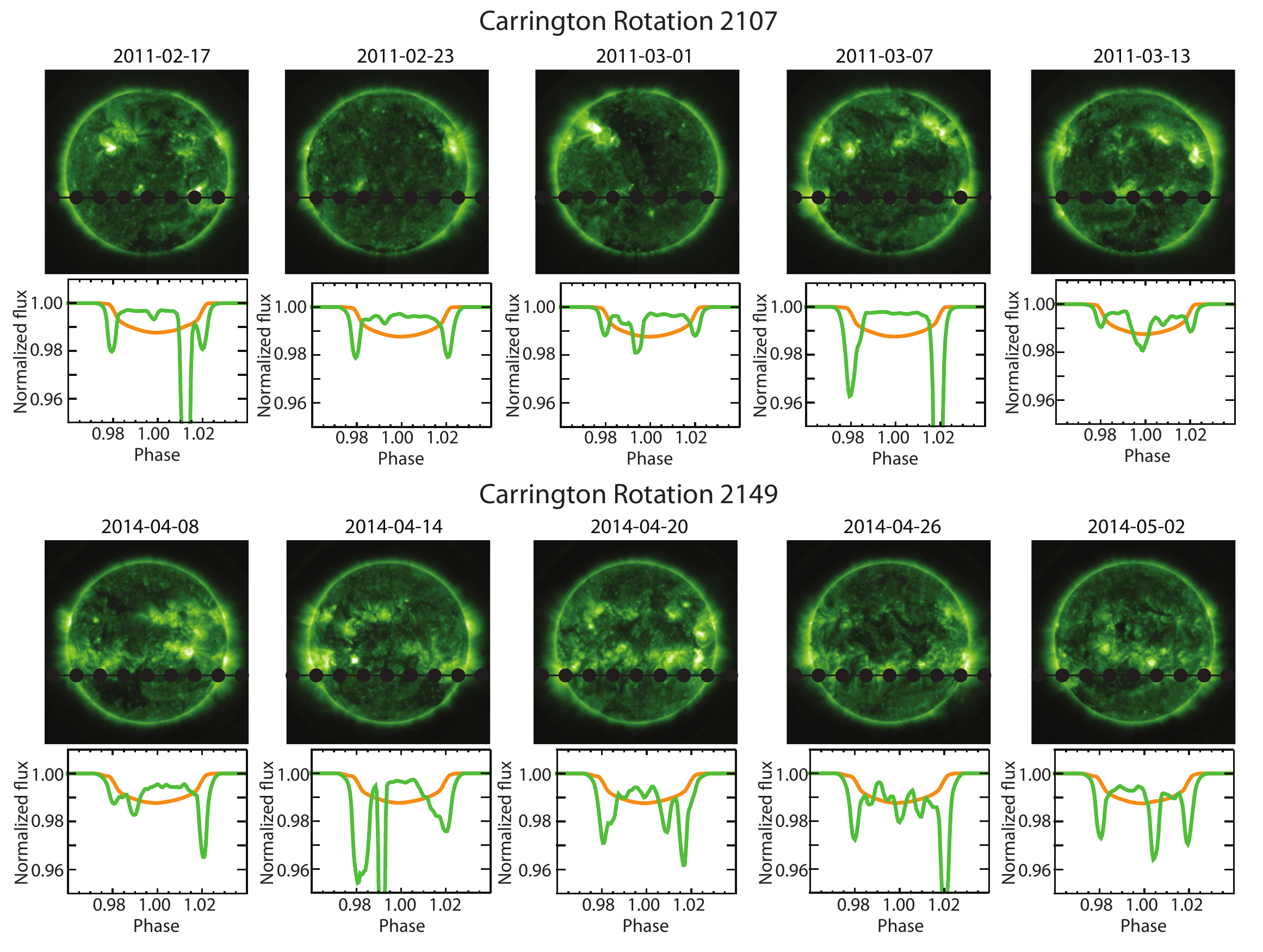} 
   \caption{ Soft X-ray (94\,\angst) images of the Sun over two complete rotations. Over plotted is the orbital path of the simulated planet ($\rp=0.1, b=-0.3$). The corresponding 94\,\angst\, transit light curves (green) and the simultaneous optical light curve (orange) is shown beneath each image. The 94\,\angst\, light curves reveal how even over short time scales the transits are highly variable due the structure and evolution of the corona and rotational modulation. As the planet transits over a bright region of the corona a large dip is registered in the 94\,\angst\, light curve but not in the optical light curve. }
   \label{fig:modulation}
\end{figure*} 
\label{sec:activity}
Figure \ref{fig:modulation} shows 94\,\angst\, images of the Sun over two Carrington rotations. Under this setup the Sun has rotated $\sim60^\circ$ between each image. The top row is for Carrington rotation 2107 (2011 February) when the Sun was in activity minimum, and the bottom row is for Carrington rotation 2149 (2014 April) when the Sun was more active (but still not at activity maximum). The corresponding 94\,\angst\, (green) and optical (orange) light curves are shown below each image for an impact parameter of $b=-0.3$. We find that even for transits separated by a few days the 94\,\angst\, light curves can be highly variable and have a different shape when compared to the previous transit. This is due to the corona being highly structured and variable over short timescales. Indeed, for more active stars, and for stars with stronger differential rotation rates the coronae may be even more dynamic than shown in Figure \ref{fig:modulation} \citep{Pevtsov:2003jx,Gibb:2014bx}. 
 
Since high energy transits are sensitive to star light from the upper atmosphere of the star including the extended corona, it may in principle be possible to use transits at these wavelengths to study the properties of the corona. The duration of transit events at these wavelengths may be longer than the corresponding optical events which are only sensitive to light from the photosphere of the star. In our simulated 94\angst\,- 335\angst\, light curves we see hints of an extended transit when the planet has transited over an active region on the stellar limb. We find that the maximum difference in phase between the 94\angst\, and optical ingress is $\sim0.005$. In principle, this effect could be used to study the extent and density profile of stellar coronae; however, very high cadence, signal-to-noise transits would be required which is currently observationally challenging at these wavelengths. 

For WASP-12b, the near-UV (2800 \AA) ingress and optical ingress timing differed by $\sim0.02$ in phase. We found no detectable timing difference between our FUV (1600 \& 1700 \AA) and optical light curves. This therefore suggests that the asymmetry in the light curve of WASP-12b is unlikely to be have been caused by stellar activity, but indeed could be caused by additional, asymmetrically distributed material surrounding the exosphere of the planet. 

\subsection{Transit depths} 
 \begin{figure} 
	\centering 
	\includegraphics[width=3in]{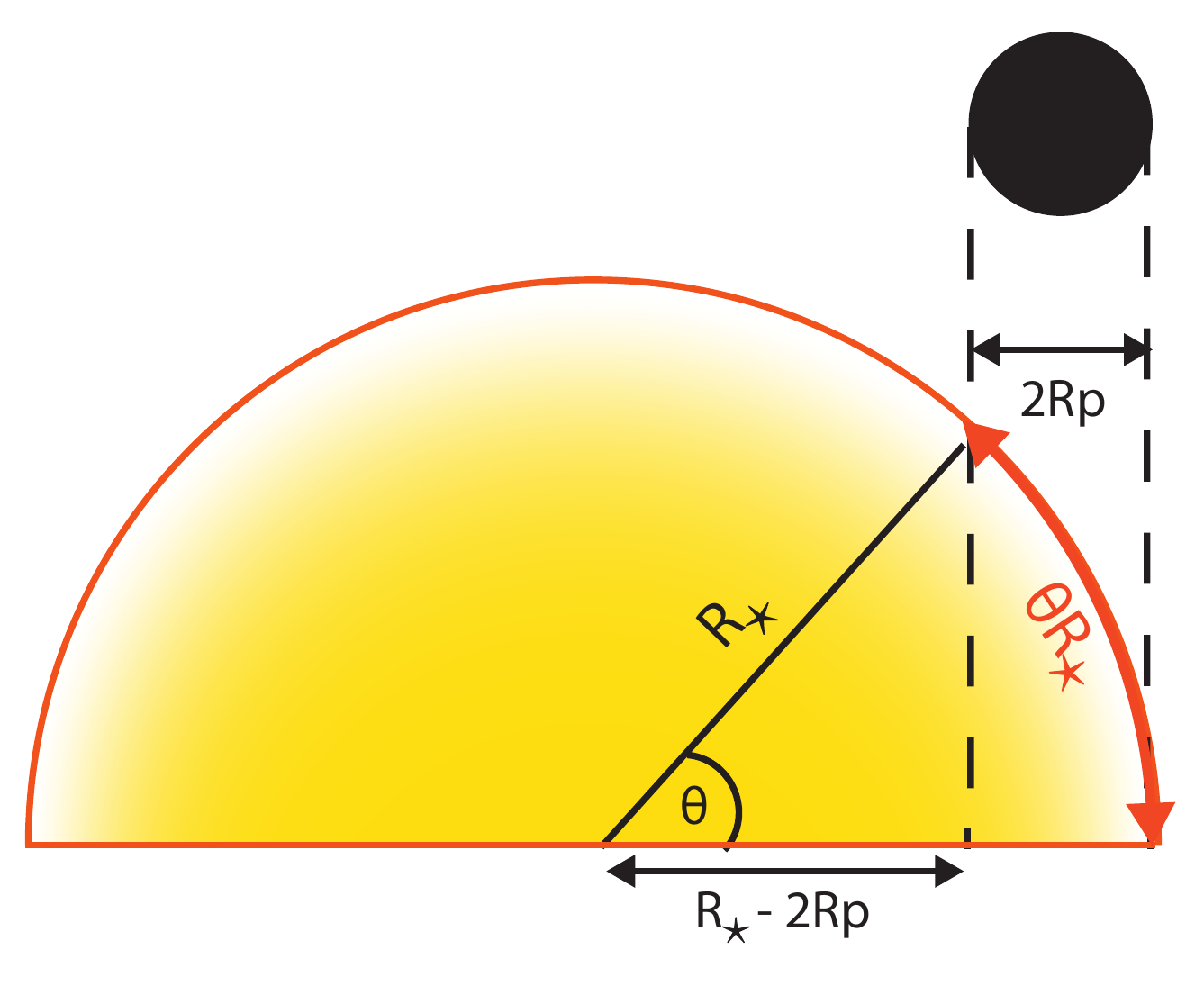}       
	\caption{Top-down schematic of a limb-brightened transit (not drawn to scale) adapted from \citet{2010ApJ...722L..75S}. The image shows how the majority of the starlight comes from the limb of the star. Under this setup, the deepest part of the transit will be at ingress and egress, rather than mid-transit for longer wavelength observations.}
	\label{fig:cartoon}   
\end{figure}
To further quantify the variability in our simulated light curves, we computed the maximum depth $\df$ for each light curve. \citet{2010ApJ...722L..75S} provide an estimate for the maximum depth of limb-brightened transits based on a model where the flux from the star is primarily concentrated at the limb rather than at disk center as is typical for optical wavelengths. Under this setup, the maximum expected transit depth (in the absence of active regions on the stellar surface) occurs at second contact of the transit.\begin{figure*} 
	\centering 
	\includegraphics[width=7in]{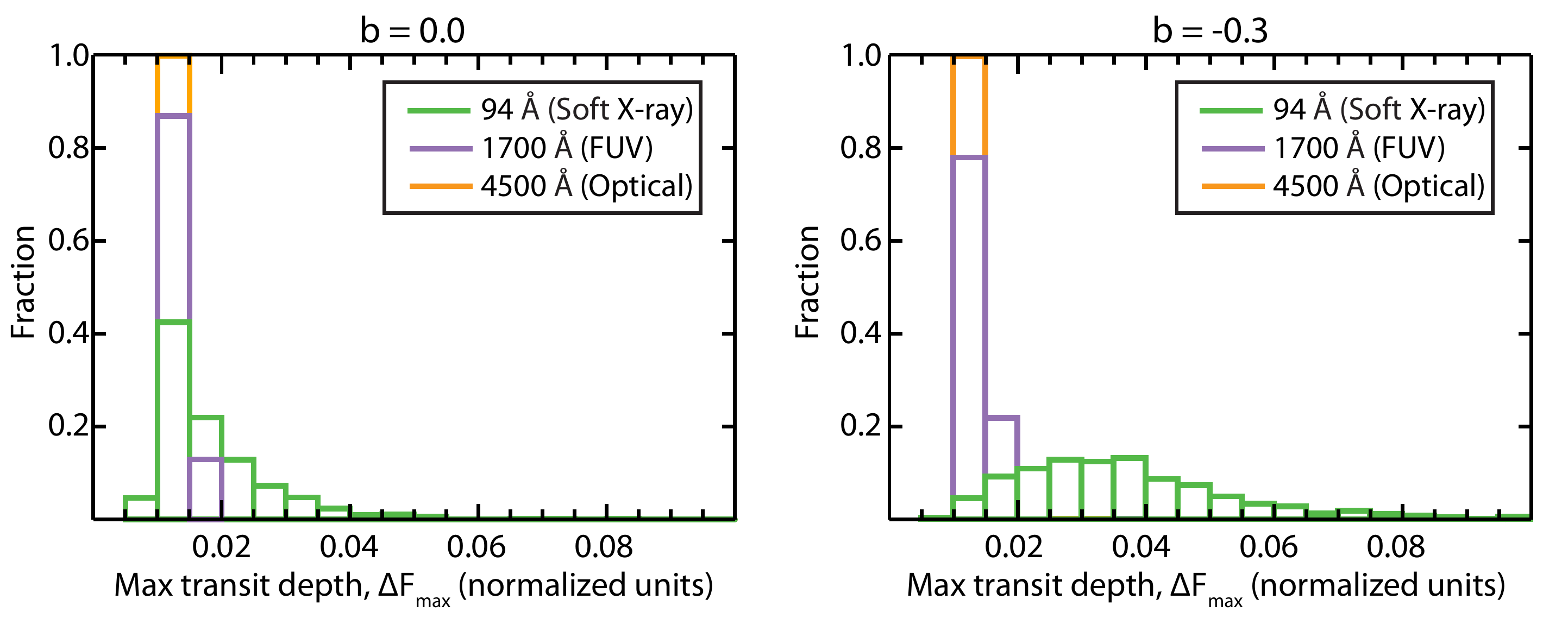}       
	\caption{Histograms of maximum transit depth, $\df$ as a function of wavelength for the 94 \AA\,(green), 1700 \AA\, (purple), and 4500 \AA\, (orange) light curves. The left panel is for an impact parameter $b=0$, the right histogram is for $b=-0.3$. In the FUV and optical we generally recover a consistent value of $\df$ (see Table \ref{tab:stats}). We find a much larger range of maximum transit depths in the soft X-ray / EUV caused by both unocculted active regions creating shallower transits and occulted active regions resulting in deeper transits. The results for the other wavelengths are shown in Table \ref{tab:stats}.}
	\label{fig:hist}
\end{figure*} 

Figure \ref{fig:cartoon} shows a top-down schematic of this setup. The length of the arc occulted by the planet is $\theta R_\star$, and $2R_p =  R_\star(1-\cos\theta) \approx \frac{1}{2}R_\star\theta^2$ for $\theta\ll1$. The area occulted by the planet can then be approximated by an ellipse with semi-major axis $\frac{1}{2}\theta R_\star$ and semi-minor axis $R_p$, implying the occulted area can be expressed as $A=\pi R_p\sqrt{R_pR_\star}$. Finally, the maximum transit depth (in normalized units) is then expressed as  
\begin{equation}  
	\frac{\Delta\textrm{F}_{\rm bright}}{\textrm{F}_{\rm bright}}= \frac{A}{2\pi \textrm{R}_\star^2} =  \frac{1}{2}\left(\frac{\textrm{R}_\textrm{p}}{\textrm{R}_\star}\right)^{3/2}.
\label{eqn:df_brightened}  
\end{equation}

For a more detailed geometric description and derivation of Equation \ref{eqn:df_brightened} we refer the reader to \citet{2010ApJ...722L..75S}. They note that for the same value of $\rp$, transits at wavelengths where the stellar disk is limb-brightened will be deeper than those where the star is limb-darkened. For a uniformly illuminated stellar disk, the maximum transit depth occurs when the entire disk of the planet is within the stellar disk (as viewed by the observer), and the normalized depth of the transit can be expressed as 
\begin{equation}
	\frac{\Delta\textrm{F}_{\rm uniform}}{\textrm{F}_{\rm uniform}}=\frac{2\pi \textrm{R}_\textrm{p}^2}{2\pi \textrm{R}_\star^2} =\left(\frac{\textrm{R}_\textrm{p}}{\textrm{R}_\star}\right)^2.
\label{eqn:df_darkened}
\end{equation}

In the presence of limb-darkening, the mid-transit depth will deviate slightly from the depth given in Equation \ref{eqn:df_darkened}. Equations \ref{eqn:df_brightened} and \ref{eqn:df_darkened} provide estimates for $\rp$; however, in practice a full transit model must be fit to the light curve to accurately derive $\rp$ (see Section \ref{sec:rp}).

Figure \ref{fig:hist} is a histogram of all values of $\df$ computed from all the simulated transit light curves for the 94 \AA, 1700 \AA, and 4500 \AA\, light curves. We find that $\df$ can deviate significantly from the predicted value of $\Delta\textrm{F}_{\rm bright}$ given in Equation \ref{eqn:df_brightened}, and that the deepest part of the transit may not be at the limb when active regions are present on the stellar disk. 

For the optical and FUV light curves we find that activity has less impact on the depth of the transit and we recovered the expected transit depth of 0.01 in $\sim80\%$ of the FUV light curves. In the optical, when the planet transits over a dark sunspot, the fractional loss in light is less and so a relatively localized, positive bump is registered in the light curve (which is excluded from the fitting). As such, we recovered the expected transit depth of 0.01 in all of our optical light curves. Figure \ref{fig:hist} reveals how the scatter in $\df$ decreases with longer wavelengths.

\begin{figure*} 
	\centering 
	\includegraphics[width=7in]{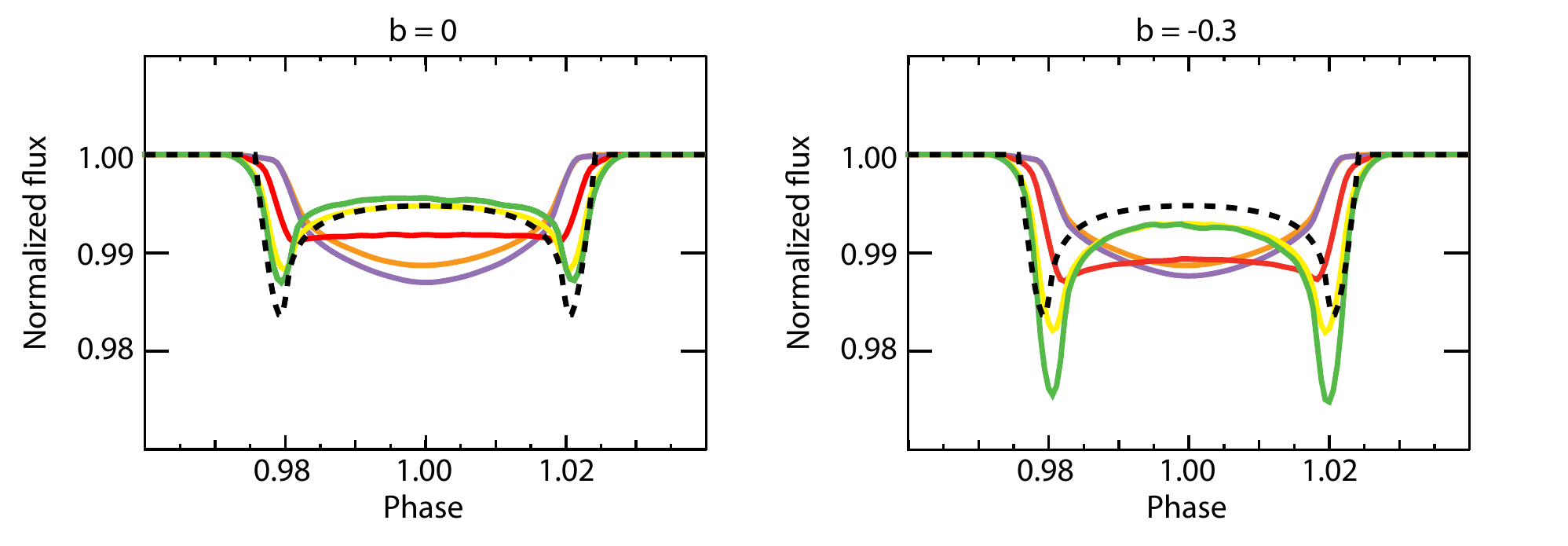}       
	\caption{Averaged light curves for our 94 \angst\, (green); 171 \angst\, (yellow); 304 \angst\, (red); 1700 \angst\, (purple); 4500 \angst\, (orange) simulated light curves. The left panel are the averaged light curves for a planet with impact parameter $b=0$ and the right panel for $b=-0.3$. The limb-brightened light curves in the left panel are shallower than those in the right panel due to the effects of unocculted bright regions on the stellar disk. The dashed line is a model limb-brightened, spot free, model transit from \citet{2010ApJ...722L..75S}.}
	\label{fig:avglcs}
\end{figure*}
The largest range in the maximum depth of the transit occurs at the shortest wavelengths (94\AA\, - 335\AA). This will have significant impact on the ability to reliably recover the radius of the planet (see Section \ref{sec:rp}). The light curves where the planet has transited over the equator of the Sun ($b=0$) are on average shallower than the light curves where the planet transits over the activity belts ($b=-0.3$). This can be seen in Figure \ref{fig:avglcs} where the average of the $\sim 1300$ simulated light curves for each wavelength in our simulations for both impact parameters is computed. These shallow transits are the direct consequence of unocculted bright regions on the stellar disk during the transit. In the normalized stellar image unocculted bright regions mean the planet occults relatively less star light during the transit, resulting in a shallower transit. For a more in-depth discussion of unocculted dark spots on the depths of transit light curves we refer the reader to Section 3 of \citet{Pont:2013ch}. Note, dark star spots on the transit depth will have the opposite effect to the bright regions present at soft X-ray and UV wavelengths.   

\subsection{Transits affected by activity}\label{sec:affected}
We quantified the percentage of our simulated light curves that show signatures of stellar activity (where the planet occults an active region) by first simulating activity-free transits using the code developed by \citet{2010ApJ...722L..75S}, itself based on the limb-darkened transit code of \citet{Mandel:2002bb}. These transit codes assume that the stellar disk is free from active regions. We computed a limb-brightened, activity-free, model transit for a hot Jupiter with $\rp=0.1$ and impact parameters $b=0$ and $b=-0.3$ (plotted as dashed black lines in Figure \ref{fig:avglcs}) to match the input to our simulations. For the UV and optical light curves we simulated limb-darkened, activity-free transits using the code developed by \citet{Mandel:2002bb}.

We then calculated the residuals for our Solar transits by subtracting the activity-free light curve from each simulated light curve to compute the residual flux and searched for deviations in the residuals caused by the planet occulting an active region. For the X-ray and UV transits, when the planet occults a bright active region a negative dip is registered in the light curve. We therefore searched the residuals for such dips. We conservatively classified a dip as being detectable if it's residual flux value was $3\times$ larger than the $1-\sigma$ uncertainty in the light curves of \citet{Poppenhaeger:2013wx}, i.e., if the residuals contained a value larger than $0.018\times$ normalized flux. 

When the planet occults an active region in the optical a positive bump is registered in the light curve. Guided by the measured uncertainties of  \citet{SanchisOjeda:2011hd} we classified a bump as being detectable if it's residual flux value was larger than $0.001\times$ normalized flux, making it detectable by \textit{Kepler}.


Columns 3 and 5 of Table \ref{tab:stats} give the fraction of light curves that show detectable signatures of the planet occulting active regions as a function of impact parameter (as defined by the criteria above). The fraction of affected light curves is heavily dependent on wavelength and geometry of the transit chord with respect to the activity belts, with $\sim15\%$ of the 94\AA\, - 335\AA\, transits where $b=0$ exhibiting detectable signatures of active region occultation. This fraction increases to $\sim70\%$ when the planet transits over the activity belts ($b=-0.3$).  $10-25\%$ of FUV light curves exhibit detectable dips caused by stellar activity depending on the orbital path of the planet with respect to the activity belts on the stellar surface. Only $\sim5\%$ of the optical light curves showed detectable bumps where the planet had transited over a dark sunspot, consistent with the findings of \citet{Llama:2012jl}. Since these bumps are localized features in the light curve they can be accounted for when computing $\rp$. 

More active stars will have a much larger fraction of the stellar surface covered with active regions, making the likelihood of the planet occulting an active region higher. This suggests that our findings are likely to serve as lower limits on the fraction of affected light curves for active stars. If these dips caused by the planet occulting active regions at short wavelengths are not accounted for when fitting transit models then an inflated value of $\rp$ will be found. Given the current time sampling and sensitivity of high energy transits it is very difficult to isolate individual bumps in the light curve.

\subsection{Recovering $\rp$ as a function of wavelength}\label{sec:rp}
We also recovered the radius ratio, $\rp$ to investigate how stellar activity impacts this measurement for both the limb-darkened and limb-brightened transits. 
To recover $\rp$ from the limb-brightened light curves we again used the full transit model from \citet{2010ApJ...722L..75S} to produce transit light curves with varying values of $\rp$ to find the best fit limb-brightened model transit. In the FUV and optical, we used the full transit model of \citet{Mandel:2002bb} with varying values of $\rp$ to find the best fit transit. We note that in the optical it is relatively straightforward to exclude regions of the light curve where the planet has occulted a dark star spot; however, at short wavelengths it is non-trivial to determine which parts of the light curve have been affected by activity and so in the soft X-ray and UV we fit the entire light curve. 
  
\begin{deluxetable*}{c|cc|cc}
\centering
\tablecaption{Summary of recovered values of $\rp$ and fraction of stellar activity impacted light curves}
\startdata
\hline
	        &     \multicolumn{2}{c}{$b=0$} & \multicolumn{2}{c}{$b=-0.3$}\\
 Wavelength &     Recovered &     Activity &Recovered &     Activity \\
 (\AA)		&	  $\rp$ &     Impacted (\%)&  $\rp$    &	 Impacted (\%) \\ 
 \hline 
94     & 0.090 $\pm$ 0.014    & 14    & 0.120 $\pm$ 0.026    & 69  \\
131    & 0.089 $\pm$ 0.012    & 10    & 0.119 $\pm$ 0.020    & 69  \\
171    & 0.093 $\pm$ 0.010    & 05    & 0.112 $\pm$ 0.015    & 40  \\
193    & 0.092 $\pm$ 0.012    & 04    & 0.118 $\pm$ 0.017    & 48  \\
211    & 0.090 $\pm$ 0.017    & 10    & 0.123 $\pm$ 0.024    & 74  \\
304    & 0.094 $\pm$ 0.008    & 11    & 0.110 $\pm$ 0.016    & 39  \\
335    & 0.085 $\pm$ 0.020    & 19    & 0.125 $\pm$ 0.031    & 64  \\
1600   & 0.113 $\pm$ 0.007    & 10    & 0.115 $\pm$ 0.009    & 24  \\
1700   & 0.118 $\pm$ 0.006    & 06    & 0.117 $\pm$ 0.007    & 16  \\
4500   & 0.100 $\pm$ 0.001    & 03    & 0.100 $\pm$ 0.001    & 05  
\enddata
\tablecomments{Column 1 is the wavelength of the observation. Columns 2 \& 3 give statistics for the simulations with $b=0$. Columns 4 \& 5 show the same statistics but for $b=-0.3$. Columns 2 \& 4 give the median value of $\rp$ and associated variability (derived from the difference between the \nth{90} and \nth{10} percentiles of the histograms shown in Figure \ref{fig:hist_rp}). Columns 3 \& 5 give the percentage of the light curves where stellar activity caused a detectable deviation from the model, activity free light curves (see Section \ref{sec:affected} for details).}
\label{tab:stats}
\end{deluxetable*}
 \begin{figure*} 
	\centering 
	\includegraphics[width=7in]{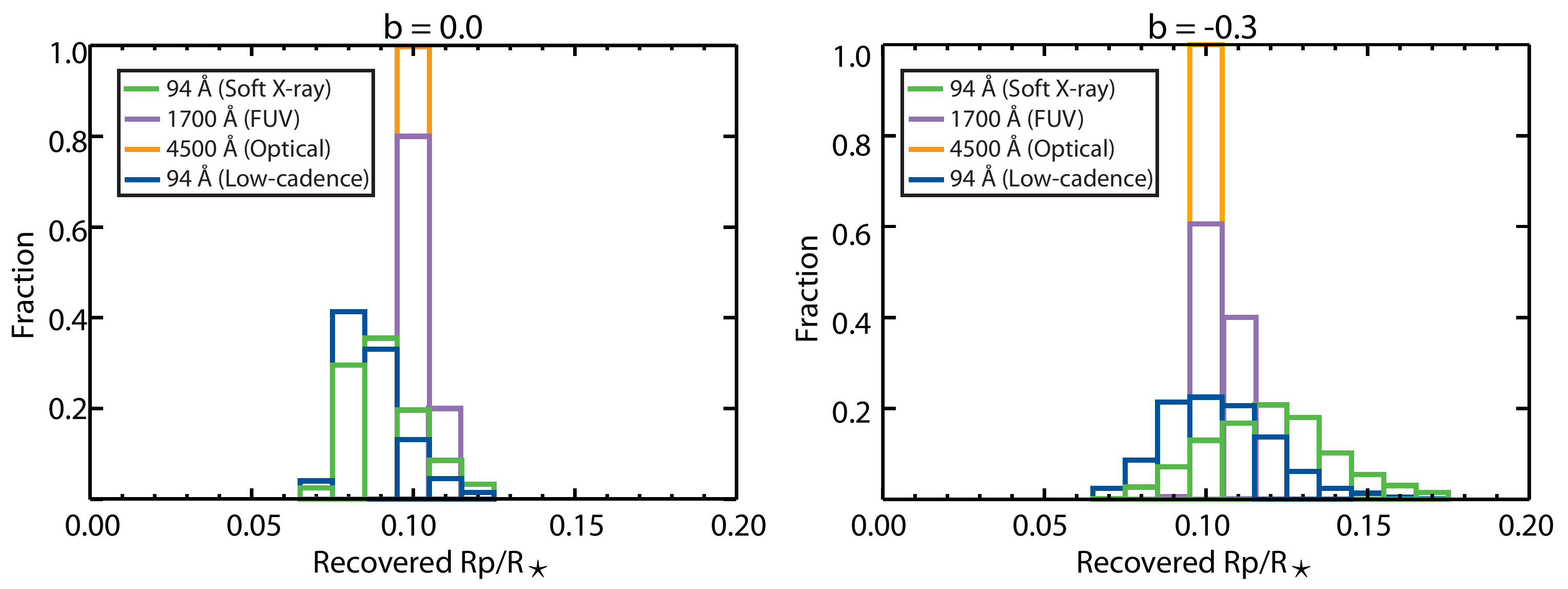}       
	\caption{Histograms displaying the recovered values of $\rp$ (see Section \ref{sec:rp} for more details) for the 94\angst\, (green), 1700\angst\, (purple), and 4500\angst\, (orange) simulated light curves. We always recover the input value of $\rp=0.1$ from the optical light curves. For the FUV light curves we find the mean value of $\rp$ to be larger than the optical value due to active regions appearing as extended bright regions. For the high energy transits (94\angst\,-335\angst) we find the values of $\rp$ to be more variable. The light curves where the planet did not transit over the activity belts ($b=0$) are affected by unocculted active regions which can result in an underestimate in the recovered value of $\rp$ of up-to 20\%. When the planet transits over the activity belts ($b=-0.3$) these light curves are much deeper, resulting in an overestimate in the recovered values of $\rp$ of up-to 50\%. The blue histogram is the recovered $\rp$ when we rebin our 94\angst\, light curves to a lower cadence (see Section \ref{sec:discussion}). }
	\label{fig:hist_rp}  
\end{figure*}
Figure \ref{fig:hist_rp} shows histograms of the recovered values of $\rp$ as a function of wavelength and impact parameter for the 94\AA, 1700\AA, and 4500\AA\, transits. Columns 2 \& 4 of Table \ref{tab:stats} give summary statistics of our recovered values of $\rp$ for all the wavelengths. We recovered the input value of $\rp=0.1$ from all of the optical light curves. The mean recovered value of $\rp$ in the FUV was $\sim20\%$ larger than the optical value, caused by the active regions appearing as extended, bright regions in the FUV rather than relatively confined, dark regions in the optical. Observations and analysis of HD 209458 have shown that the UV (3000\angst - 3900\angst) transit depth is some 0.030\% deeper than in the optical (4500\angst - 5500\angst) \citep{Ballester:2007cc,Sing:2008ce,LecavelierDesEtangs:2008ef}. \citet{LecavelierDesEtangs:2008ef} attribute this additional absorption to Rayleigh scattering in the atmosphere of the planet. For such small changes in transit depth it is therefore imperative that the effects of unocculted active regions (which can also change the transit depth) are accounted for in UV transit light curve analysis.
\begin{figure*} 
	\centering 
	\includegraphics[width=7in]{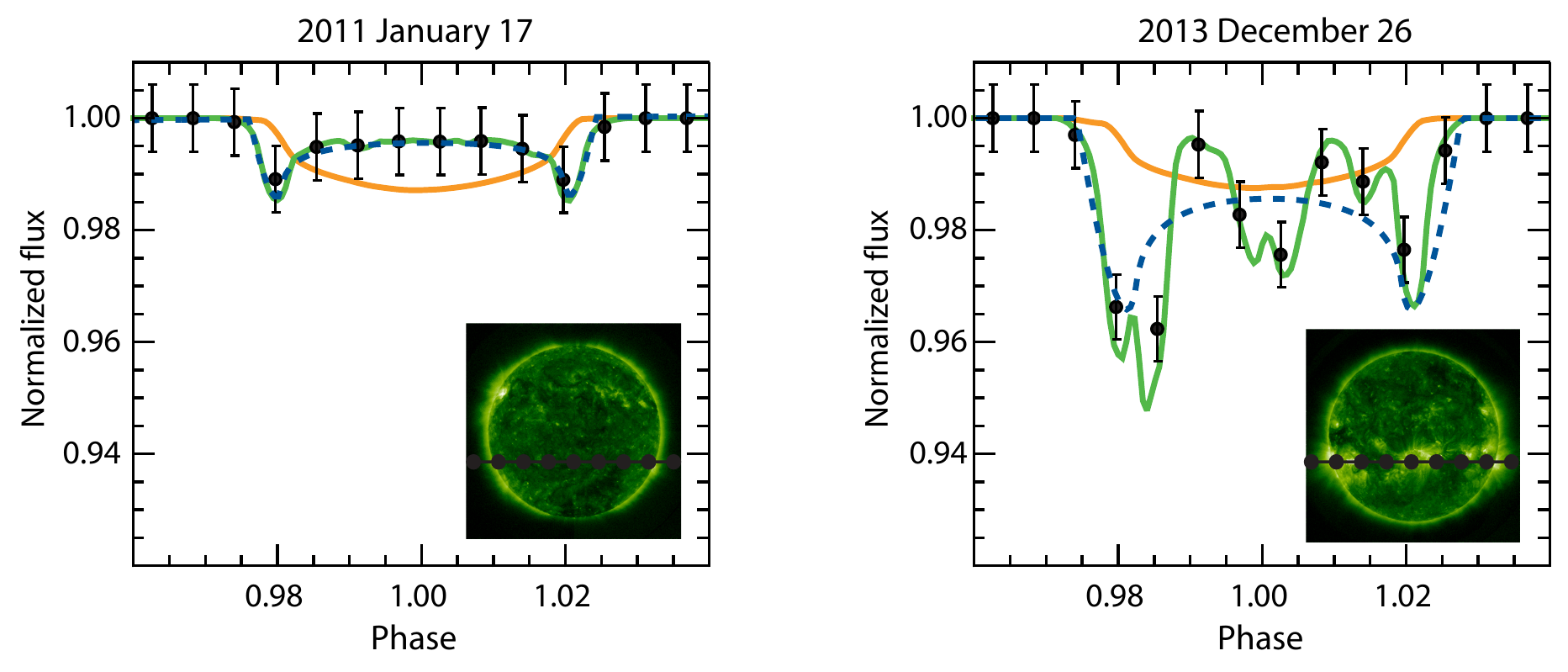}       
	\caption{The optical light curve (orange) and 94 \AA\, (green) for nearly-quiescent (left) and activity impacted (right) cases. The best fit transit light curve to the X-ray light curve is shown as the dashed blue line. In the lower-left of each panel is the corresponding 94 \AA\, image. The black points are the result of rebinning the 94 \AA\, light curve and adding a similar level of noise to the levels in the observations of \citet{Poppenhaeger:2013wx}. We find that the maximum depth of the 94 \AA\,low-cadence light curves is typically shallower than for the high-cadence light curves; however, fewer data points make it more challenging to distinguish between an increased transit being caused by stellar activity, or by an extended planetary atmosphere.}
	\label{fig:lowcad}
\end{figure*}

There was a much wider spread in the recovered radius values at the shortest wavelengths. The radius of the planet was underestimated by up-to 25\% in the light curves where the planet did not transit over the activity belts ($b=0$). This underestimate is due to a shallower transit caused by the presence of unocculted bright regions on the stellar disk. When the planet did transit over the activity belts ($b=-0.3$), the recovered value of $\rp$ was found to be over-estimated by up-to 50\%, caused by the planet transiting over bright regions resulting in a deeper transit. This can be seen in Figure \ref{fig:lowcad} where we have plotted the best fit X-ray light curve (dashed blue line) for two light curves. Again, for more active stars these values will serve as lower limits.  
 
\section{Implications for X-ray transits}\label{sec:discussion}
X-ray (1\angst\, - 63\angst) transit observations of HD 189733b have revealed a transit depth up-to 4.8 times deeper than the optical transit depth which has been attributed to the planet having an extended atmosphere \citep{Poppenhaeger:2013wx}. These X-ray light curves have a mid-transit depth shallower than the depth in the optical. In our simulated short wavelength (94 \angst\, - 335 \angst) transits, where $\rp$ is kept constant, only those light curves unaffected by stellar activity have mid-transit depths shallower than the optical depth (see the right panel in Figure \ref{fig:transits}). This suggests that either HD 189733b may indeed have an extended atmosphere in X-rays, or that during those observations the planet was transiting over X-ray bright regions that covered a large fraction of the transit path. Both of these scenarios would result in an X-ray mid-transit depth deeper than the corresponding optical transit depth. This second scenario is not unreasonable given that HD 189733 is a more active star than the Sun. Indeed, Zeeman-Doppler Imaging of HD 189733 has revealed the star has a large-scale magnetic field strength ten times stronger than that of the Sun \citep{Fares:2010hq}. 

More X-ray transits will therefore help determine which scenario is correct. We find that for large active regions, a bump in the optical light curve coincides with large dips in the high energy light curves, suggesting that optical light curves could potentially be used to rule out stellar activity effects on deep X-ray light curves. Therefore, simultaneous optical light curves will also aide the interpretation.

Due to the difficulty in obtaining X-ray observations, the light curves of \citet{Poppenhaeger:2013wx} are at a lower cadence than those simulated here. In our high-cadence observations the signatures of the planet transiting over an active region can be resolved as localized dips in the light curve; however, this may not always be the case for lower cadence observations. The black points in Figure \ref{fig:lowcad} show the effects of rebinning two 94\angst\, light curves to the same cadence of the X-ray transits of \citet{Poppenhaeger:2013wx} and adding similar sized error bars. These rebinned light curves make it much more difficult to detect the signatures of the planet transiting over active regions. The right panel of Figure \ref{fig:lowcad} shows how the localized dips in the high cadence light curve caused by the planet occulting an active region are less defined at lower cadence. If the planet occults larger (or more) active regions than in this case then the depth of the X-ray transit will be consistently lower than the optical transit depth. This may lead to the false conclusion that the planet has an inflated atmosphere in X-rays.

To quantify the effects of using lower cadence observations we repeated the analysis carried out in Section \ref{sec:rp} and derived the value of $\rp$ using the low cadence 94 \AA\, light curves. The dark blue histogram in Figure \ref{fig:hist} gives the results for this analysis. We find a reasonable agreement between the recovered values of $\rp$ for the light curves when the planet is not occulting the activity belts ($b=0$). When the planet is occulting the activity belts ($b=-0.3$), the low cadence light curves produce smaller values of $\rp$ when compared to the full cadence observations. This is caused by the active regions being localized regions, and so integrating for longer during the observations reduces their impact on the light curve; however, the disadvantage of having fewer data points is that it is less obvious whether the increased depth in the light curve is caused by the planet or stellar activity.

\section{Conclusions}
In this paper we used resolved images of the Solar disk to simulate transit light curves at multiple wavelengths ranging from soft X-rays/EUV to optical. By simulating a transiting planet with two impact parameters we investigated the importance of both occulted and unocculted active regions on transit light curves.

At short wavelengths, up-to 70\% of the simulated transit light curves where the planet occulted the activity belts on the Sun showed signatures of stellar activity that would result in the derived value of $\rp$ deviating from the real value. This value decreased to $\sim20\%$ for the light curves when the planet's orbital path did not occult the activity belts. At longer wavelengths, the effects of stellar activity were not as prevalent, with $20\%$ of our simulated FUV light curves where the planet occulted the activity belts showing signatures of stellar activity. In the optical, we found that for stars with activity levels similar to the Sun, only 5\% of light curves will have bumps caused by the planet transiting over star spots agreeing with the findings of \citet{Llama:2012jl}.

We also investigated the effects stellar activity would have on recovering the radius ratio of the planet and the host star, $\rp$. If the planet transits over the activity belts on the star, then the mean value of the planet's radius as derived from soft X-ray and EUV light curves can be overestimated by 20\%; however, during times of high activity the radius can be overestimated by up-to 50\%. When the planet did not occult the activity belts, the effects of unocculted active regions resulted in the mean value of the planet's radius at short wavelengths being underestimated by 10\%. Again, during periods of high activity this underestimate increased to $\sim25\%$.

The effects were similar in the FUV for which the radius of the planet could be over-estimated by $\sim15\%$. In the optical we recovered the input value of $\rp=0.1$ in all our simulated light curves since bumps in the light curve can be accounted for when deriving $\rp$. 

We compared our light curves to other transit observations at short wavelengths. We found that transits at short wavelengths showed hints of an extended duration than the corresponding optical transits. This is due to these observations being sensitive to light from both the photosphere and extended corona, suggesting high-energy transits could be used to study the density and structure of stellar coronae. Unlike the limb-brightened transits, the FUV (1600 \angst\, \& 1700 \angst) light curves exhibited no signatures of an early ingress when compared to the optical. This suggests that the observed early-ingress in the near-UV (2800 \angst) transit of WASP-12 is unlikely to be caused by stellar activity alone; rather, it strengthens the idea that WASP-12b has additional absorbing material surrounding the exosphere of the planet. 
 
By rebinning our high-cadence soft X-ray (94 \angst) light curves to a similar cadence as the X-ray data of HD 189733b taken by \citet{Poppenhaeger:2013wx}, we demonstrated that it is difficult to distinguish between a deeper X-ray transit being caused by stellar activity, or by an extended planetary atmosphere. Higher time sampling will make identifying parts of the light curve where the planet has occulted an active region easier. Simultaneous optical data may also aide in revealing whether the planet is transiting over active regions and help distinguish between these two scenarios.  

We conclude that it is very important that when attempting to determine planetary properties using transits at high energy wavelengths, multiple observations are obtained with long baselines, and that simultaneous optical data (both during the transit and out-of-transit) may help decouple planetary properties from stellar variability.  
 
 \vspace{-0.3in}  
\acknowledgements 
This work is supported by NASA Origins of the Solar System grant No. NNX13AH79G. The authors thank A.\,A. Vidotto (Geneva) \& L. Fletcher (Glasgow) for providing helpful comments and suggestions. We would also like to thank the anonymous referee for his/her feedback and thought provoking questions on the manuscript.
   
\bibliography{sdo}  
\end{document}